\documentclass[twocolumn,journal]{IEEEtran}

\ifCLASSINFOpdf
  \usepackage[pdftex]{graphicx}
  \DeclareGraphicsExtensions{.pdf,.jpeg,.png}
\else
  \usepackage[dvips]{graphicx}
  \DeclareGraphicsExtensions{.eps}
\fi
\usepackage{epstopdf} %MFM to Allow eps figures

\usepackage{amsmath}
\usepackage{amssymb}
\usepackage{tabu}

\usepackage{comment}
\usepackage{multicol}
\usepackage{amsthm}
\usepackage{float}
\usepackage{color}
\usepackage{balance} 

\usepackage{cite}
\newcommand{\Tr}{\text{Tr}}
\newcommand{\diag}{\text{diag}}
\def\BState{\State\hskip-\ALG@thistlm}
\makeatother

\usepackage[belowskip=-10pt,aboveskip=0pt]{caption}
\begin{document}
	\title{\Large Securing OFDM-Based NOMA SWIPT Systems}
%\author{\small Author 1 and Author 2
%}
\author{ 
Ahmed~Badawy, \IEEEmembership{Member,~IEEE,} and Ahmed El Shafie, \IEEEmembership{Senior Member,~IEEE} 
      % <-this % stops a space
\thanks{A. Badawy is with with Politecnico di Torino, DET., Turin, Italy (e-mail: ahmed.badawy@polito.it).} \thanks{A. El Shafie is with Qualcomm Tech. Inc, San Diego, CA 92121 USA (e-mail: ahmed.salahelshafie@gmail.com).}}

% <-this % stops a space
%\thanks{Part of this work was presented at 2013 IEEE Pacific Rim Conference on Communications, Computers and Signal Processing (IEEE PacRim 2013) and The 9th IEEE International Conference on Wireless and Mobile Computing, Networking and Communications (WiMob 2013)}
%\thanks{This research work is funded by Qatar National Research Fund, QNRF (a member of Qatar Foundation, QF) under grant number NPRP 7-923-2-344. The statements made herein are the sole responsibility of the authors.}
%}% <-this % stops a space

	\maketitle
	\begin{abstract}
		%\boldmath
In this paper, we present a physical-layer security scheme that exploits artificial noise (AN) to secure the downlink legitimate communications and transfer energy to nodes operating under non-cooperative non-orthogonal multiple-access (NOMA) scenario. The nodes employ a joint time-switching and power-switching scheme to maximize the harvested energy. We provide necessary analysis and derivations for the optimization parameters and find the optimized transmission parameters that maximize the minimum secrecy rate among users while meeting constraints on minimum transferred energy and outage probabilities at the nodes through an exhaustive grid-based search. Our analysis and simulations prove the feasibility of securing the communication among NOMA nodes, while transferring energy and meeting outage probability constraints. 
	\end{abstract}
\section{Introduction}
%Research in non-orthogonal multiple-access (NOMA) scheme has gained significant attraction recently. It has also been a potential multiple-access scheme for future new radio (NR) fifth generation (5G) systems. Power-domain NOMA explores users' channel conditions to improve spectrum efficiency. Moreover, nodes with energy harvesting (EH) capabilities can transfer received radio frequency (RF) signals into direct current (DC) to power the nodes for future signal processing and data transmission tasks. Under simultaneous wireless information and power transfer (SWIPT), both information and energy could be extracted from the same RF signal at the same time. 

%\subsection{Related Work}
In \cite{Liu2016}, the authors proposed cooperative non-orthogonal multiple-access (NOMA) scheme where only the near user is an energy harvesting (EH) node that relays information to the far user, however, the work does not take into account securing the communication between the multiple nodes. The work in \cite{Tang19} optimized energy efficiency in non-cooperative NOMA under power budget and data rate constraints without exploiting physical-layer security schemes \cite{Badawy_IWCMC}. Hybrid simultaneous wireless information and power transfer (SWIPT) schemes were proposed in \cite{Shafie17} and \cite{8943160} under the consideration of the harvested energy at the receiving node while securing the legitimate transmissions, however, both works do not take NOMA and/or multiple-access scenarios into consideration. Moreover, \cite{Gao17} investigated physical layer security in large scale NOMA networks, without considering EH at the nodes. %investigates physical layer security under different relaying protocols for cooperative NOMA scenario without considering EH at the nodes.}

\subsection*{Contributions}	
In this paper, we consider a system operating under non-cooperative NOMA scenario which employs OFDM transmissions. Without knowledge of the eavesdroppers' instantaneous channel state information (CSI) since the eavesdroppers are assumed to be passive, the base station (BS) aims at securing communication between itself and legitimate users with the help of artificial noise (AN). In addition, the scheme transfers energy to the legitimate users. We provide necessary analysis and derivations for the achievable secrecy rates, energy transfer rates and outage probability at the legitimate non-cooperative far NOMA users. Several design parameters play different roles in maximizing the secrecy rates while meeting constraints on minimum outage probability and maximum transferred energy, such as cyclic-prefix (CP) length, time-switching (TS) factor between energy transfer and data transmission, power-splitting factor (PS), power-allocation factor between data signal and AN signal, PS factor between NOMA users under a desired outage probability at the legitimate users, and a desired average energy transfer rate constraints to maximize the secrecy rate between BS and legitimate NOMA users. We find the optimized values for these parameters through an exhaustive grid-based search.
	
\subsection*{Notation}	
Lower and  upper case bold letters denote vectors and matrices, respectively. $(\cdot)^\dagger$, $(\cdot)^*$, $\diag\{\cdot\}$,  $\Tr\{\cdot\}$ and $||\cdot||_F$ denote transpose, Hermitian, diagonal, trace and the Frobenius norm of a matrix, respectively. $\boldsymbol{\mathrm{F}}$ is the FFT matrix, $\boldsymbol{\mathrm{F^*}}$ is the IFFT matrix. $\boldsymbol{\mathrm{I}}_N$ is the identity matrix with dimension $N \times N$. $\boldsymbol{\mathrm{0}}_{N_1 \times N_2}$ is the zero matrix with dimensions $N_1 \times N_2$.  $\mathbb{R}^{N_1 \times N_2}$ is the set of real numbers with dimensions $N_1 \times N_2$ and $\mathbb{C}^{N_1 \times N_2}$ is the set of complex numbers with dimensions $N_1 \times N_2$. $\mathbb{E}\{.\}$ denotes the expectation. $\bar{\rho} = 1 - \rho$. $[\cdot]^+  = \max\{\cdot, 0\}$. $\mathcal{CN}(\cdot,\star)$ is a Gaussian distribution with `$\cdot$' mean and `$\star$' variance.
	\section{System Model}
		\begin{figure}
		\centering
		\includegraphics[width=1.65in]{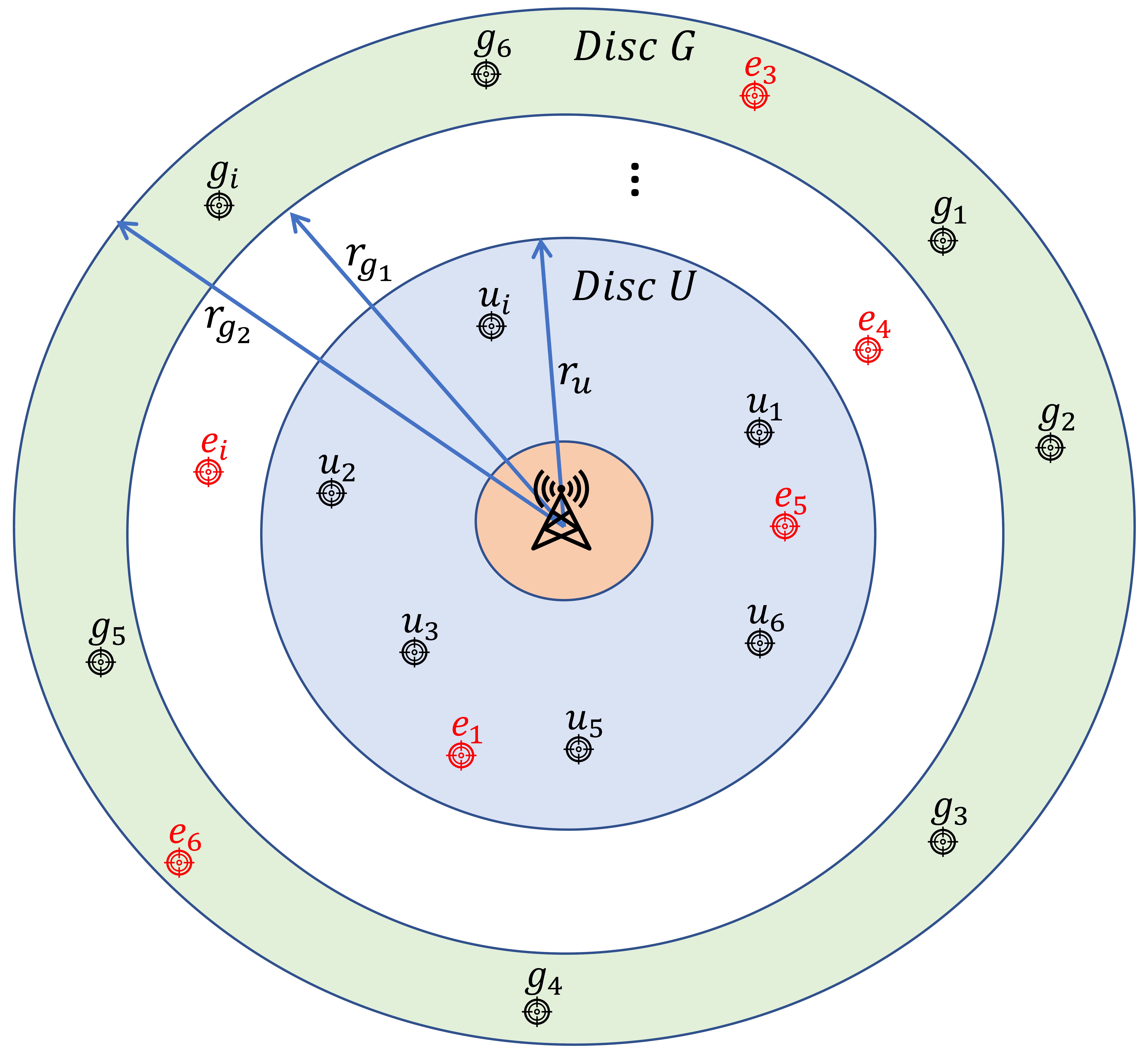}
		\caption{Network Topology.}
		\label{fig1}
	\end{figure}
In our adopted system model, we assume a BS trying to secure communication between itself and multiple legitimate users in the presence of passive eavesdroppers (Eves), denoted as $e$, as depicted in Fig. \ref{fig1}. To this end, the users are categorized into two groups according to the radius of the disc they are located in and the BS is assumed to lie in the center of the discs. Without loss of generality, we assume two discs only, inner Disc $U$ with radius $r_u$ and Disc $G$ with inner radius $r_{g_1}$ and outer radius $r_{g_2}$. We assume that the spatial topology for both legitimate users and eavesdroppers is modeled according to Poisson point processes (PPP). The legitimate users are assumed to have reliable power supplies. It is also assumed that the legitimate users are EH nodes. Since the energy required for transmission is much higher than that required for processing of the received samples \cite{Nasir13}, to save energy at the near user, a non-cooperative NOMA transmission scheme is adopted between the BS and the legitimate users. The eavesdroppers are assumed to be located at least half wavelength apart from any legitimate user.% The communication is carried out in an OFDM framework, where the total available BW is divided into a number of orthogonal sub-channels. All nodes are assumed to be using single antenna for transmission and reception.  
%with density $\lambda$ and denoted by $\zeta$. 
	
It is assumed that communication is carried out under OFDM scenario. We consider an OFDM symbol with total duration time $T$, which is composed of the CP time duration, denoted by $T_{\rm cp}$, and the time duration of the useful data signal, denoted by $T_s$. To this end, $T = N_t/f_s = N_{\rm cp}/f_s + N_s / f_s$, where $N_t = N_{\rm cp} + N_s$ is the total number of samples within one OFDM symbol, $N_{\rm cp}$ is the number of the samples within the CP, $N_s$ is the number of samples of the transmitted data signal, and $f_s$ is the sampling frequency. Under OFDM scheme, the available bandwidth is divided into $N_s$ orthogonal sub-channels.
	
The wireless channels between BS and users are modeled as block-fading channels, which implies that the channel coefficients do not vary during the channel coherence time. The thermal noise at the receiving nodes is modeled as additive white Gaussian noise with zero mean and power $\sigma^2$.
\section{Scheme Design and Analysis}
The BS uses $N_s$-point inverse fast Fourier transform (IFFT) to convert the frequency-domain sub-channels into the time-domain. In addition, it appends a CP of length $N_{\rm cp}$ to the beginning of each OFDM symbol to prevent the inter-OFDM-symbol-interference across adjacent OFDM symbols. Along with the transmitted OFDM symbol, the BS transmits a time-domain AN signal by exploiting the temporal dimensions gained by the presence of the CP. As will be shown later, the AN is designed such that it is canceled out at the legitimate receivers prior to signal demodulation and retrieval of the transmitted data message. The BS splits its total available transmit power, $P_t$, between the data and AN signals. The power split factor is denoted by $0\le \rho \le 1$. The allocated power to the data signal is $\rho P_t$ and the allocated power to AN signal is $\bar{\rho} P_t$. 
  
BS transmits signals to two NOMA users selected from Discs $U$ and $G$, which are $u_i$ and $g_i$, respectively along with AN signal. Moving forward, we will drop the subscript for the users and use $u$ and $g$ to denote the users. The transmitted signal from BS is  
\begin{equation}
\boldsymbol{\mathrm{x}} =\sqrt{\rho P_t}\boldsymbol{\mathrm{E_{\rm cp}F^*}}\boldsymbol{\mathrm{ S p}} + \sqrt{\bar{\rho}P_t} \boldsymbol{\mathrm{K w}}, 
\end{equation}
\noindent where $\boldsymbol{\mathrm{x}} \in \mathbb{\mathbb{C}} ^{N_t \times 1}$, $\boldsymbol{\mathrm{E_{\rm cp}}} \in \mathbb{\mathbb{R}} ^{N_t \times N_s}$ is the CP insertion matrix, $\boldsymbol{\mathrm{S}}$ $ ~\in \mathbb{C} ^{N_s \times 2}$ contains the messages to be sent to the two NOMA users, $\boldsymbol{\mathrm{p}}= [\sqrt{\theta_u} \hspace{0.1in} \sqrt{\theta_g}]^\dagger$ is the vector that contains the assigned power factors for Users $u$ and $g$, respectively, $\boldsymbol{\mathrm{K}} \in \mathbb{C} ^{N_t \times N_{\rm cp}}$ is the AN precoding matrix and $\boldsymbol{\mathrm{w}} \in \mathbb{C} ^{N_{\rm cp} \times 1}$ is the AN vector.

The entire CP is used in analog domain to transfer energy to two NOMA Users, $u$ and $g$. Prior to processing and in time domain the rest of the OFDM symbol is divided into two portions. During the first portion, which has a duration $T_u = N_u/f_s$ and $T_g= N_g/f_s$ at Users $u$ and $g$, respectively, where $N_u$ and $N_g$ are the number of samples in the first portions at Users $u$ and $g$, respectively, the users use PS scheme, where a fraction of the received samples power is used for EH. For the remaining portion of the OFDM symbol, no PS is used and, hence, the entire second portion is used for data processing. 

The received signal at User $u$ is
\begin{align}
y_{i} = \begin{cases} c_u\sqrt{\beta_u \rho P_t}\left[\boldsymbol{\mathrm{ H_{{t}} E_{\rm cp} F^* S p }}\right]_i+  c_u \sqrt{\beta_u \bar{\rho}P_t}\left[\boldsymbol{\rm{H_{t} K w }}\right]_i\\ \quad   +n_{u_{i}} 
%\\ \hspace{1.1in} \text{for} \hspace{0.1in} i=N_{\rm cp} + 1,\ldots, N_{\rm cp} + N_{o}\\
\hspace{.75in} \text{for} \hspace{0.1in} i= N_{\rm cp}+1,\ldots, N_{u} +N_{\rm cp}\\
c_u \sqrt{\rho P_t}\left[\boldsymbol{\mathrm{ H_{{t}} E_{\rm cp} F^* S p }}\right]_i+ \sqrt{\bar{\rho} P_t}\hspace{0.in} c_u \left[\boldsymbol{\rm{H_{t} K w }}\right]_i \\ \quad  + n_{u_{i}},  
\hspace{0.7in} \text{for} \hspace{0.1in} i=N_{\rm cp}+N_u + 1,\ldots, N_s\\
%\hspace{1.75in} i=N_{o}+1,\ldots, N_t
\end{cases}
\end{align}
\noindent where $\beta_u$ is the power split factor during $T_u$, $c_u = \frac{1}{\sqrt{1+ d_u^\alpha}}$, $d_u$ is the distance between BS and $u$, $\alpha$ is the path loss exponent, $\boldsymbol{\mathrm{ H_t}}$ is the Toeplitz channel matrix between BS and $u$ with the impulse response of the channel as its first column, and $\bold{n}_u$ is noise at receiver $u$. Ditto, the received signal at User $g$ is
\begin{equation}
z_{i} = \begin{cases} 
c_g\sqrt{\beta_g \rho P_t}\left[\boldsymbol{\mathrm{ G_{{t}} E_{\rm cp} F^* S p }}\right]_i+  c_g \sqrt{\beta_g \bar{\rho}P_t}\left[\boldsymbol{\rm{G_{t} K w }}\right]_i \\ 
\quad+ n_{g_{i}}  \hspace{.75in} \text{for} \hspace{0.1in} i= N_{\rm cp}+1,\ldots, N_{g}+ N_{\rm cp} \\
c_g \sqrt{\rho P_t}\left[\boldsymbol{\mathrm{ G_{{t}} E_{\rm cp} F^* S p }}\right]_i+ c_g \sqrt{\bar{\rho} P_t} \left[\boldsymbol{\rm{G_{t} K w }}\right]_i \\
\quad  + n_{g_{i}},  
\hspace{.7in} \text{for} \hspace{0.1in} i=N_{\rm cp}+N_g + 1,\ldots, N_s
\end{cases}
\end{equation}
\noindent where $\beta_g$ is the power split factor during $T_g$,  $c_g = \frac{1}{\sqrt{1+ d_g^\alpha}}$, $d_g$ is the distance between BS and $g$, $\boldsymbol{\mathrm{ G_t}}$ is the Toeplitz channel matrix between BS and $g$ and $\bold{n}_g$ is the AWGN at receiver $g$.   

Before processing the symbols at the receiver side, equalization is needed to remove the impact of power split. The received samples up to sample $N_{\rm cp}+N_u$ are now
\begin{align}
\tilde{y}_{i}&=y_{i}/\sqrt{\beta_u} \nonumber
\\&= c_u  \sqrt{\rho P_t} \left[\boldsymbol{\mathrm{ H_{{t}} E_{\rm cp} F^* S p }}\right]_i+  c_u \sqrt{\bar{\rho} P_t}  \left[\boldsymbol{\rm{H_{t} K w }}\right]_i \nonumber\\
&\quad \quad \quad \quad +  n_{u_{i}}/\sqrt{\beta_u}. \label{y_tilda} %\nonumber
%\\ \hspace{1.95in} \text{for} \hspace{0.1in} i= 1,\ldots, N_{o}
\end{align}
\noindent Hence, the received vector at User $u$ is $\boldsymbol{\mathrm{y}} = [\tilde{y}_{1}, \tilde{y}_{2}, \ldots, \tilde{y}_{N_u}, y_{N_u+ N_{\rm cp}+ 1}, \ldots,  y_{N_s}]^\dagger$. The CP is removed from the received signal at $u$ by multiplying by the CP removal matrix $\boldsymbol{\mathrm{\Phi}} \in \mathbb{\mathbb{R}} ^{N_s\times N_t}$. The outcome is then fed to the FFT block which yields
\begin{align}
\boldsymbol{\mathrm{ Fy}} \!=\! c_u  \sqrt{\rho P_t}\boldsymbol{\mathrm{ F \Phi H_t E_{\rm cp} F^* S p}} &\!+\! c_u  \sqrt{\bar{\rho} P_t}\boldsymbol{\mathrm{ F \Phi H_t K w}}   \!+\!\bold{F}\boldsymbol{\mathrm{n_u}}, \label{eqn_y_fft}
\end{align}
\noindent where $\boldsymbol{\mathrm{n_u}} \in \mathbb{C} ^ {N_s \times 1}$ is the AWGN vector after CP removal. We will denote part of the first term in the left hand side of (\ref{eqn_y_fft}) by $\boldsymbol{\mathrm{H = F \Phi H_t E_{\rm cp} F^*}}$. Note that $\boldsymbol{\mathrm{H}} \in \mathbb{C} ^{N_s \times N_s}$ is a diagonal matrix, whose diagonal elements are the frequency domain channel coefficients between the BS and $u$. Similar steps are followed at $g$, which yields
\begin{align}
\boldsymbol{\mathrm{ Fz}} = c_g  \sqrt{\rho P_t}\boldsymbol{\mathrm{ G S p}} +c_g  \sqrt{\bar{\rho} P_t} \boldsymbol{\mathrm{ F \Phi G_t K w}} + \bold{F} \boldsymbol{\mathrm{n_g}}, \label{eqn_g_fft}
\end{align}
\noindent with $\boldsymbol{\mathrm{G = F \Phi G_t E_{\rm cp} F^*}}$ and $\boldsymbol{\mathrm{n_g}} \in \mathbb{C} ^ {N_s \times 1}$ is the AWGN vector after CP removal for User $g$. 

To remove the impact of AN at both legitimate users, the AN precoding matrix should satisfy
\begin{align}
\boldsymbol{\mathrm{\Phi \left(H_t + G_t\right) K}}= \boldsymbol{\mathrm{0}}_{N_s \times N_{\rm cp}} \label{eqn_null}
\end{align}

\noindent Since the rank of the matrix $\boldsymbol{\mathrm{\Phi \left(H_t + G_t\right)}}$ is $N_s$ and its number of columns is $N_t$, (\ref{eqn_null}) has a non-trivial solution and correspondingly $\boldsymbol{\mathrm{\Phi \left(H_t + G_t\right)}}$ has a non-trivial null space. 

Note that due to the equalization step applied in (\ref{y_tilda}), the noise variance for the first $N_u$ and $N_g$ samples increases. The average noise variance across all samples for Users $u$ and $g$ can be given by
\begin{align}
\small
&\sigma_{a,u}^2 = \sigma^2 \left(\frac{ N_s-N_u +\frac{N_u}{\beta_u} }{N_s}\right), \label{sigma_ave1}\\
&\sigma_{a,g}^2 = \sigma^2 \left(\frac{ N_s-N_g +\frac{N_g}{\beta_g} }{N_s}\right).\label{sigma_ave2}
\end{align} 
\subsection{Secrecy Rates}
Both $u$ and $g$ are not impacted by AN signals by design. Under NOMA scheme, $u$ decodes and removes the message sent to User $g$. The achievable rate at $u$ is
\begin{align}
R_u = \frac{1}{N_t} {\log_2 \det{\left(\boldsymbol{\mathrm I}_{N_s}+{\frac{\rho \theta_u P_t}{N_t}\boldsymbol{\mathrm{H H^* \left(F\Lambda_u F^*\right)^{-1}}}}\right)} }, \label{R_u}
\end{align}
\noindent where $\boldsymbol{\mathrm \Lambda}_u = \frac{\sigma^2}{c_u^2} \diag \{ \underbrace{{1}/{\beta_u}, {1}/{\beta_u} , \ldots, {1}/{\beta_u}}_{N_{u}}, 1, 1, \ldots, 1\} \in \mathbb{R} ^{N_s \times N_s}$. 
Under NOMA, User $g$ treats the signal assigned to User $u$ as noise and does not remove it from its received signal. Hence, the achievable rate at $g$ is
\begin{align}
R_g = \frac{1}{N_t} {\log_2 \det{\left(\boldsymbol{\mathrm I}_{N_s}+{\frac{\rho \theta_g P_t}{N_t}\boldsymbol{\mathrm{G G^* \left(F\Lambda_g F^*\right)^{-1}}}}\right)} }, \label{R_g}
\end{align}
\noindent where $\boldsymbol{\mathrm \Lambda}_g \in \mathbb{R} ^{N_s \times N_s}$ is given by $\boldsymbol{\mathrm \Lambda}_g \!=\! \left(\frac{\sigma^2}{c_g^2} \boldsymbol{\mathrm I}_{N_s} \!+\!\frac{ \rho \theta_u P_t \boldsymbol{\mathrm G} \boldsymbol{\mathrm G} ^ *}{N_t}\right)  \diag  \{\! \underbrace{{1}/{\beta_g},  \cdots,  {1}/{\beta_g}}_{N_{g}}, 1,  \cdots, 1\!\}.$

There are few cases for the the achievable rate at Eve. First if Eve is interested in decoding either of the messages sent to $u$ or $g$. In this case, Eve considers the signal that she is not interested in as noise. If Eve is interested in decoding $u$'s message only, Eve's achievable rate in this case is given by 
\begin{equation}
\begin{split}
&R_{E,u} = \frac{1}{N_t} \log_2 \det \bigg(
\boldsymbol{\mathrm I}_{N_s}+ \\& 
\frac{\rho \theta_u P_t}{N_t}\boldsymbol{\mathrm{V V^*}} 
\left(
\frac{\bar{\rho} P_t}{N_{\rm cp}}\boldsymbol{\mathrm {D D^*}} + 
\frac{\sigma^2 }{c_e^2} \boldsymbol{\mathrm I}_{N_s} +\frac{\rho \theta_g P_t \boldsymbol{\mathrm{VV^*}} }{N_t}%\right) \boldsymbol{\mathrm I_{N_s}}
\right) ^{-1}
\bigg),
\end{split}
\end{equation}
\noindent where $\boldsymbol{\mathrm{V}}$ is a diagonal matrix with the frequency domain channel coefficients between BS and Eve as its diagonal and  $\boldsymbol{\mathrm{D = F \Phi V_t K}}$, with $\boldsymbol{\mathrm{V_t}}$ being the Toeplitz channel matrix between BS and Eve. Note that the term $\frac{\bar{\rho} P_t}{N_{\rm cp}}\boldsymbol{\mathrm {D D^*}}$, which is the AN covariance matrix, represents the impact of AN at Eve. If Eve is interested in decoding $g$'s message only, Eve's achievable rate in this case is given by 
\begin{equation}
\begin{split}
&R_{E,g} = \frac{1}{N_t} \log_2 \det \bigg(
\boldsymbol{\mathrm I}_{N_s}+ \\& 
\frac{\rho \theta_g P_t}{N_t}\boldsymbol{\mathrm{V V^*}} 
\left(
\frac{\bar{\rho} P_t}{N_{\rm cp}}\boldsymbol{\mathrm {D D^*}} + 
\frac{\sigma^2 }{c_e^2}\boldsymbol{\mathrm I}_{N_s} +\frac{\rho \theta_u P_t \boldsymbol{\mathrm{V V^*}}}{N_t}%\right) \boldsymbol{\mathrm I_{N_s}}
\right) ^{-1}
\bigg).
\end{split}
\end{equation}
If Eve is interested in decoding both signals and employs a joint-typicality receiver, the secrecy rate can be given by
\begin{equation}
\begin{split}
R_E = &\frac{1}{N_t} \log_2 \det \bigg(
\boldsymbol{\mathrm I}_{N_s}+ \\&
\frac{\rho P_t}{N_t}\boldsymbol{\mathrm{V V^*}} 
\left(
\frac{\bar{\rho} P_t}{N_{\rm cp}}\boldsymbol{\mathrm {D D^*}} + 
\frac{\sigma^2}{c_e^2} \boldsymbol{\mathrm I_{N_s}}
\right) ^{-1}
\bigg).
\end{split}
\end{equation}
The secure transmission rate between BS and $u$ and $g$ following the previous cases can now be given by
\begin{align}
 &R_{s,u} \leq [R_u-R_{E,u}]^+, \label{Rs1_u}\\
 &R_{s,g} \leq [R_g-R_{E,g}]^+. \label{Rs1_g}
\end{align}
\noindent The sum secrecy rate is upper-bounded by \cite{Tekin08}
\begin{align}
   R_{s,u} + R_{s,g} \leq \left[R_u + R_g - R_E\right]^+. \label{R_sum}
\end{align}
The rate pair ($R_{s,u} ,~ R_{s,g}$) should satisfy (\ref{Rs1_u}), (\ref{Rs1_g}) and (\ref{R_sum}).
%%%====================================================================================
%%%====================================================================================
%%%====================================================================================
%%%====================================================================================
%%%====================================================================================
%%%====================================================================================
%%%====================================================================================
%%%====================================================================================
%%%====================================================================================
%%%====================================================================================
\subsection{Outage Probability}
The outage probability for $u$, $P_{o,u}$, comprises the probability that $u$ cannot detect the signal sent to $g$ and the probability that $u$ can detect the message sent to $g$ but cannot detect the message sent to itself. This is given as
\begin{align}
\small
P_{o,u} =& \Pr\left\{c_u^2||\boldsymbol{\mathrm{H}}||_F^2 < \vartheta_{\{g,u\}} \right\} \nonumber \\ 
&+ \Pr\left\{c_u^2||\boldsymbol{\mathrm{H}}||_F^2 > \vartheta_{\{g,u\}},  \vartheta_{\{g,u\}} < \vartheta_{\{u,u\}} \right\},
\end{align}
\noindent where $\vartheta_{\{g,u\}} = \frac{\delta_1 \sigma_{a,u}^2}{ \rho P_t \left( \theta_g  -    \theta_u  \delta_1 \right)}$, $\vartheta_{\{u,u\}} = \frac{\delta_2 \sigma_{a,u}^2}{ \rho \theta_u  P_t}$, with $\delta_1$ satisfies $\theta_g - \delta_1 \theta_u > 0$ \cite{Ding2014} and $\delta_2 = 2^{(2\delta_3)} - 1$, where $\delta_3$ is the desired rate at which $u$ can detect the message sent to itself. Hence, with the help of the results provided in \cite{Liu2016} under the assumption that the spatial topology for the users is modeled through PPP, the approximate outage probability at User $u$ per sub-channel can be given by

%$\rho$ is equivalent to $\theta$ in shafie's paper. $p_{i1}$ in el kashlan paper equivalent to $\theta_g$ . Far User $A$ in el kashlan paper equivalent to $g$

\begin{equation}
\small
P_{o,u} \approx
\begin{cases}  
\begin{split} &\frac{\pi}{2L} \sum_{l=1}^{L}\sqrt{1-\cos^2{\left(\frac{2 l - 1}{2L} \pi\right)}}
 \times \\ &\left( 1 - \exp\left(- n_l \vartheta_{\{g,u\}} \right)\right) 
  \times  \left(1+ \cos{\left(\frac{2 l - 1}{2L} \pi\right)}\right), \\& 
  \hspace{1.72in} \text{for} \quad \vartheta_{\{g,u\}} \geq \vartheta_{\{u,u\}}
\end{split} \\
1 \hspace{1.65in} \text{for} \quad \vartheta_{\{g,u\}} < \vartheta_{\{u,u\}}
\end{cases}
\end{equation}
\noindent where $n_l = 1 + \left(\frac{r_u}{2}\left(1+\cos{\left(\frac{2 l - 1}{2L} \pi\right)}\right)\right)^\alpha$, and $L$ is a design parameter used for complexity-accuracy tradeoff. The outage probability for $g$, denoted by $P_{o,g}$, can be given by
\begin{align}
P_{o,g} = \Pr\left\{c_g^2||\boldsymbol{\mathrm{G}}||_F^2 < \vartheta_{\{g,g\}}\right\},
\end{align}
\noindent where $\vartheta_{\{g,g\}} = \frac{\delta_1 \sigma_{a,g}^2}{\rho P_t \left(\theta_g - \theta_u \delta_1\right)}$. We extend the derivation provided in \cite{Liu2016} for cooperative NOMA to our non-cooperative NOMA model and, hence, the approximate outage probability for User $g$ per sub-channel can be given by
%\begin{equation}
%\begin{split} 
%P_{o,g}& \approx \frac{\pi \left(r_{g_2} - r_{g_1}\right)}{2M\left(r_{g_2}+r_{g_1}\right)} \sum_{m=1}^{M}\sqrt{1-\cos^2{\left(\frac{2 m - 1}{2M} \pi\right)}}\\
%&\times \left( 1 - \exp\left(- \left( 1 + n_m^\alpha  \right)\vartheta_{\{g,g\}} \right)\right) \\
%& \times  \left(1+ \cos{\left(\frac{2 m - 1}{2M} \pi\right)}\right), 
%\end{split} 
%\end{equation}
\begin{equation}
\begin{split} 
P_{o,g}& \approx \frac{\pi}{M\left(r_{g_2}+r_{g_1}\right)} \sum_{m=1}^{M}\sqrt{1-\cos^2{\left(\frac{2 m - 1}{2M} \pi\right)}}\\
&n_m \times \left( 1 - \exp\left(- \left( 1 + n_m^\alpha  \right)\vartheta_{\{g,g\}} \right)\right) \\
%& \times  \left(r_{g_1}+\frac{\Delta_{r_g}}{2} \left(1+ \cos{\left(\frac{2 m - 1}{2M} \pi\right)}\right)\right), 
\end{split} 
\end{equation}
\noindent where $n_m = \frac{\Delta_{r_g}}{2}\left(1+ \cos{\left(\frac{2 m - 1}{2M} \pi\right)}\right) +r_{g_1}$, $\Delta_{r_g} = r_{g_2} - r_{g_1}$ and $M$ is a design parameter used for complexity-accuracy tradeoff. 
%%%====================================================================================
%%%====================================================================================
%%%====================================================================================
%%%====================================================================================
%%%====================================================================================
%%%====================================================================================
%%%====================================================================================
%%%====================================================================================
%%%====================================================================================
%%%====================================================================================
\subsection{Harvested Energy}The energy harvested at both users comprises of two portions. The first portion is from the CP signal, which is
\begin{equation}
\boldsymbol{\mathrm{\upsilon }}_u=c_u \boldsymbol{\mathrm{A_{\rm cp} H_t }}\left(\sqrt{\rho P_t}\boldsymbol{\mathrm{E_{\rm cp}F^*}}\boldsymbol{\mathrm{ S p}} + \sqrt{\bar{\rho} P_t}\boldsymbol{\mathrm{K w}}\right),
\end{equation}
\noindent for User $u$ and defined similarly for User $g$ as
\begin{equation}
\boldsymbol{\mathrm{\upsilon }}_g= c_g\boldsymbol{\mathrm{A_{\rm cp} G_t}}\left(\sqrt{\rho P_t}\boldsymbol{\mathrm{E_{\rm cp}F^*}}\boldsymbol{\mathrm{ S p}} + \sqrt{\bar{\rho} P_t}\boldsymbol{\mathrm{K w}}\right),
\end{equation}
\noindent where $\boldsymbol{\mathrm{A_{\rm cp}}} = \left[\boldsymbol{\mathrm{I}}_{N_{\rm cp}} \quad \boldsymbol{\mathrm{0}}_{N_{\rm cp} \times N_s}\right] \in \mathbb{R} ^{N_{\rm cp} \times N}$ is the CP extraction matrix. The harvested energy during CP is then
\begin{align}
\small
E_{1,u} &= \nonumber\\ & \eta  P_t T_{\rm cp} c_u^2 \Tr \bigg\{\boldsymbol{\mathrm{A_{\rm cp} H_t}}\bigg( \frac{\rho\boldsymbol{\mathrm{E_{\rm cp} E_{\rm cp}^*}}}{N_t} +  \frac{\bar{\rho}\boldsymbol{\mathrm{K K^*}}}{N_{\rm cp}}\bigg)\boldsymbol{\mathrm{H_t^* A^*_{\rm cp}}} 
\bigg\},
\end{align}
\begin{align}
\small
E_{1,g} &= \nonumber \\& \eta  P_t T_{\rm cp} c_g^2 \Tr \left\{\boldsymbol{\mathrm{A_{\rm cp} G_t}}\left( \frac{\rho\boldsymbol{\mathrm{E_{\rm cp} E_{\rm cp}^*}}}{N_t} + \frac{\bar{\rho}\boldsymbol{\mathrm{K K^*}}}{N_{\rm cp}}\right)\boldsymbol{\mathrm{G_t^* A^*_{\rm cp}}}
\right\},
\end{align}
\noindent where $0\le \eta \le 1$ is the efficiency factor of the RF energy conversion process at the energy harvester circuit. The second portion of the energy is harvested from $N_u$ and $N_g$ samples after CP and can be given by
%\begin{align}
%E_{2,u} =  \frac{\gamma \rho P_t N_u T_s c_u \bar{\mu}}{N} \Tr\left\{ %\boldsymbol{\mathrm{B}}_{N_u}\boldsymbol{\mathrm{B}}_{N_u}^*\right\}
%\end{align}
%\begin{equation}
%E_{2,u} =  \frac{\eta c_u^2 \rho \theta_u \bar{\beta_u} P_t N_u T_s  %}{N_t} \Tr\left\{ \boldsymbol{\mathrm{A_{N_u}\Phi H_t E_{\rm cp}}}  \left( %\boldsymbol{\mathrm{A_{N_u}\Phi H_t E_{\rm cp}}} \right)^*\right\}
%\end{equation}
%\begin{equation}
%E_{2,g} =  \frac{\eta c_g^2\rho \theta_g \bar{\beta_g} P_t N_g T_s  }{N} \Tr\left\{ %\boldsymbol{\mathrm{A_{N_g}\Phi G_t E_{\rm cp}}}  \left( \boldsymbol{\mathrm{A_{N_g}\Phi G_t E_{\rm cp}}} %\right)^*\right\}
%\end{equation}

\begin{equation}
\small
E_{2,u} =  \frac{\eta c_u^2 \rho  \bar{\beta_u} P_t N_u T_s  }{N_t} \Tr\bigg\{ \boldsymbol{\mathrm{A_{N_u}\Phi H_t E_{\rm cp}}}  \left( \boldsymbol{\mathrm{A_{N_u}\Phi H_t E_{\rm cp}}} \right)^*\bigg\}
\end{equation}
\begin{equation}
\small
E_{2,g} =  \frac{\eta c_g^2\rho  \bar{\beta_g} P_t N_g T_s  }{N_t} \Tr\bigg\{ \boldsymbol{\mathrm{A_{N_g}\Phi G_t E_{\rm cp}}}  \left( \boldsymbol{\mathrm{A_{N_g}\Phi G_t E_{\rm cp}}} \right)^*\bigg\}
\end{equation}
where $\boldsymbol{\mathrm{A_{N_u}}} = \left[\boldsymbol{\mathrm{I}}_{N_{u}} \quad \boldsymbol{\mathrm{0}}_{N_{u} \times (N_s - N_u)}\right] \in \mathbb{R} ^{N_{u} \times N_s}$ is the $N_u$ samples extraction matrix and $\boldsymbol{\mathrm{A_{N_g}}} = \left[\boldsymbol{\mathrm{I}}_{N_{g}} \quad \boldsymbol{\mathrm{0}}_{N_{g} \times (N_s - N_g)}\right] \in \mathbb{R} ^{N_{g} \times N_s}$ is the $N_g$ samples extraction matrix.
%\begin{align}
%E_{2,u} =  \frac{\gamma \rho P_t N_u T_s c_g \bar{\mu}}{N} \Tr\left\{ \boldsymbol{\mathrm{B}}_{N_g}\boldsymbol{\mathrm{B}}_{N_g}^*\right\}
%\end{align}
The total energy at $u$ and $g$ can be given by
\begin{align}
E_u = E_{1,u} + E_{2,u}, \ E_g = E_{1,g} + E_{2,g}.
\end{align}
%%%====================================================================================
%%%====================================================================================
%%%====================================================================================
%%%====================================================================================
%%%====================================================================================\\
%%%====================================================================================
%%%====================================================================================
%%%====================================================================================
%%%====================================================================================
%%%====================================================================================
\subsection{Optimization Problem}
We optimize the minimum average secrecy rate of Users $u$ and $g$ according to
\begin{equation}
\begin{split}
\max_{\beta_u, \beta_g, \rho, \theta_u, \theta_g, N_u, N_g, N_{\rm cp}}: %\mathbb{E} \left\{ \min \left\{R_{s,u} \quad R_{s,g} \right] \right\}\\
\min \left\{\mathbb{E} \left\{R_{s,u} \right\}, \mathbb{E} \left\{R_{s,g} \right\}\right\},\\
\text{s.t.} \quad  \mathbb{E} \left\{E_u\right\} \geq \mu_u, \quad \mathbb{E} \left\{E_g\right\} \geq \mu_g, \\
P_{o,u}  \leq \epsilon_u, \quad P_{o,g} \leq \epsilon_g, \\
0 \leq \beta_u, \beta_g, \rho, \theta_u, \theta_g \leq 1, \\
N_u, N_g \in \left\{0, 1, 2, \ldots, N_s \right\}, \\
%N_{\rm cp} \in \left\{\tau, \tau+1 , \ldots, N_s\right\}
\tau \leq N_{\rm cp}/f_s \leq N_s/f_s,
\end{split} \label{eqn_opt}
\end{equation}
\noindent where $\mu_u \geq 0$ and $\mu_g \geq$ are the desired average energy harvested rates at Users $u$ and $g$, respectively, $0\leq \delta_u \leq 1$ and $0\leq \delta_g \leq 1$ are the desired outage probability at Users $u$ and $g$, respectively, and $\tau = \max \{\tau_u,\tau_g\}$ with $\tau_u$, $\tau_g$ are the delay spreads between BS and Users $u$ and $g$, respectively. 

\noindent \textbf{Some Remarks on Eqn. (\ref{eqn_opt}):}
There are some tradeoffs in (\ref{eqn_opt}) as follows \begin{itemize}
	%\item The minimum length of the CP is the maximum of the delay spread between BS and $u$ and $g$ to ensure no interblock interference occurs. The transmission data rate is maximized when the CP length is minimized, i.e., $N_{\rm cp}T_s = \tau$. However, the longer the CP, the higher the harvested energy at both users.
	\item As $N_u$ and $N_g$ increase, the harvested energy at Users $u$ and $g$ increase. However, as can be seen from  Eqns. (\ref{sigma_ave1}) and (\ref{sigma_ave2}), $\sigma_{a,u}^2, \sigma_{a,g} ^2 \geq \sigma^2$, i.e., the average noise variance increases as $N_u$ and $N_g$ increase, which implies lower SNR at users. Each user can adjust its split samples, i.e., $N_u$ or $N_g$, based on its needs. Similar remarks can be stated for $\beta_u$ and $\beta_g$.
	\item As $\rho$ increases, SNR at the users increases. However, this means that less power is dedicated to AN which will also increase the SNR at the eavesdropper and, hence, reduce the secrecy rate.
	\item The NOMA power split factor should satisfy $\theta_u < \theta_g$ and $\theta_u +\theta_g = 1$ while as $\theta_u$ increases, the SNR and hence the rate at User $u$ increases, however, the SINR and hence the rate at User $g$ decreases as can be seen from Eqns. (\ref{R_u}) and (\ref{R_g}). 
	%\item we might play with which discs to select from ... here we assumed U and G only. what if we assume more? this will have impact on $c_u$ and $c_g$ More like user pair selection problem, ??????????
\end{itemize}
Due to the non-convexity of the objective function and the constraints, the optimization problem in (\ref{eqn_opt}) is not convex and hence it can solved offline using Matlab’s  fmincon  function  or  multi-dimensional grid-based search over the optimization variables and then the feasible set optimal values get communicated to both NOMA Users $u$ and $g$ before operation.  The optimization problem is solved through an exhaustive grid-based search at the BS, which is also known as brute force solution\footnote{The associated complexity can be reduced by using alternative computationally-efficient methods such as the interior-point methods \cite[Chapter 11]{resp8} which are used by the fmincon function in Matlab. Devising a more efficient method to solve the non-convex optimization problem in (\ref{eqn_opt}) is out of the scope of this paper. We shall investigate that in our future work.}. Under the transmission parameters such as $N_t$ and $N_{cp}$, the BS calculates $R_{s,u}$, $R_{s,g}$ ,$P_{o,u}$, $P_{o,g}$, for all feasible sets of the optimization parameters. The BS then selects the values that maximize the minimum secrecy rates between itself and $u$ and $g$, while meeting constraints on $E_u$, $E_g$, $P_{o,u}$ and $P_{o,g}$. The calculated optimal values should remain the same as long as the system’s average parameters including required $\mathbb{E} \left\{E_u\right\}$, $\mathbb{E} \left\{E_g\right\}$, $\mathbb{E} \{P_{o,u}\}$, $\mathbb{E} \{P_{o,g}\}$ and average channel  gains remain the same. The optimization problem does not require knowledge of eavesdroppers' instantaneous CSI. Only statistics are required to be estimated. 
%%%====================================================================================
%%%====================================================================================
%%%====================================================================================
%%%====================================================================================
%%%====================================================================================\\
%%%====================================================================================
%%%====================================================================================
%%%====================================================================================
%%%====================================================================================
%%%====================================================================================
\section{Simulation Results}
We conduct our OFDM-based NOMA SWIPT system simulation using Rayleigh fading channel realizations that follow $\mathcal{CN}(0,1)$. Independent channel realizations are generated for each user and for the eavesdropper. The rest of the simulation parameters are $\sigma^2 = 1$ Watt, $N_t = 64$, $N_{\rm cp} = 16$, $f_s = 2$ MHz,  $r_u = 8$ meters, $r_{g_1} = 10$ meters, $r_{g_2} = 14$ meters, $d_u = 4$ meters, $d_g = 12$ meters, $d_e = 10$ meters, $\eta = 0.75$, and $\alpha = 2$. The calculated ranges for SNR at $g$ and $u$ across all variations of $N_g$, $\beta_g$, $\theta_g$, $\theta_u$, $\rho$ are $\in[0:13]$ dB and $\in[27:45]$ dB, respectively. Due to limited available space, we present more results for $g$ since it is the far user operating under non-cooperative NOMA. Moreover, we simulate for fixed $N_u = 16$ and $\beta_u = 0.5$, however, similar conclusions for varying these two parameters are analogous to varying $N_g$ and $\beta_g$.
The optimized values are estimated through an exhaustive grid-based search operation at the BS. We present these values at the end of the section. In the following figures, we show the impact of varying the optimization parameters on the performance of the system.

\begin{figure}
		\centering
		\includegraphics[width=3.8in]{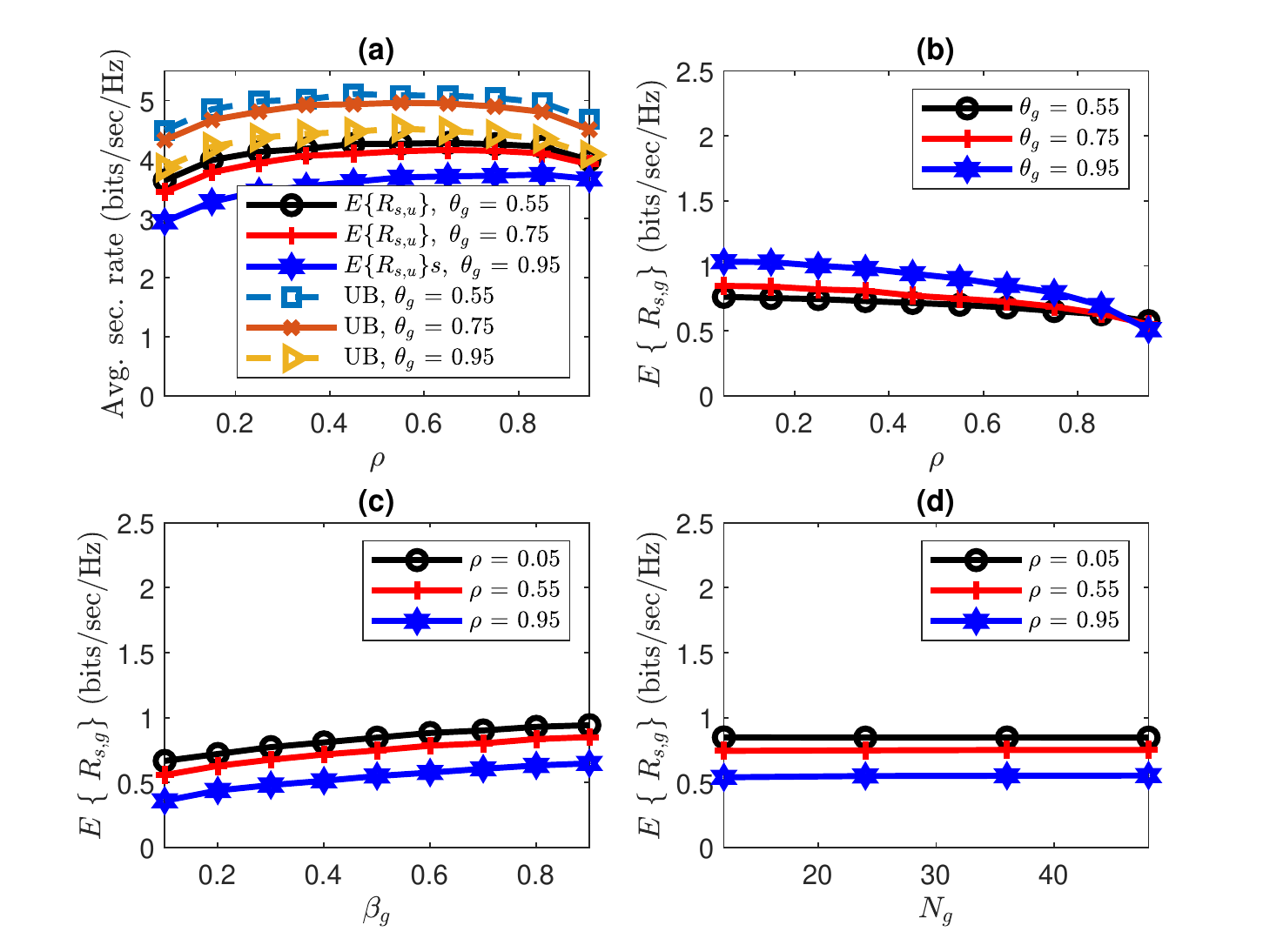}
		\caption{Average secrecy rates: (a) $\mathbb{E}\{R_{s,u}\}$ and UBs, (b) $\mathbb{E}\{R_{s,g}\}$ vs. $\rho$, (c) $\mathbb{E}\{R_{s,g}\}$ vs. $\beta_g$ and (d) $\mathbb{E}\{R_{s,g}\}$ vs. $N_g$.}
		\label{sec_rates}
\end{figure}

\begin{figure}
		\centering
		\includegraphics[width=3.8in]{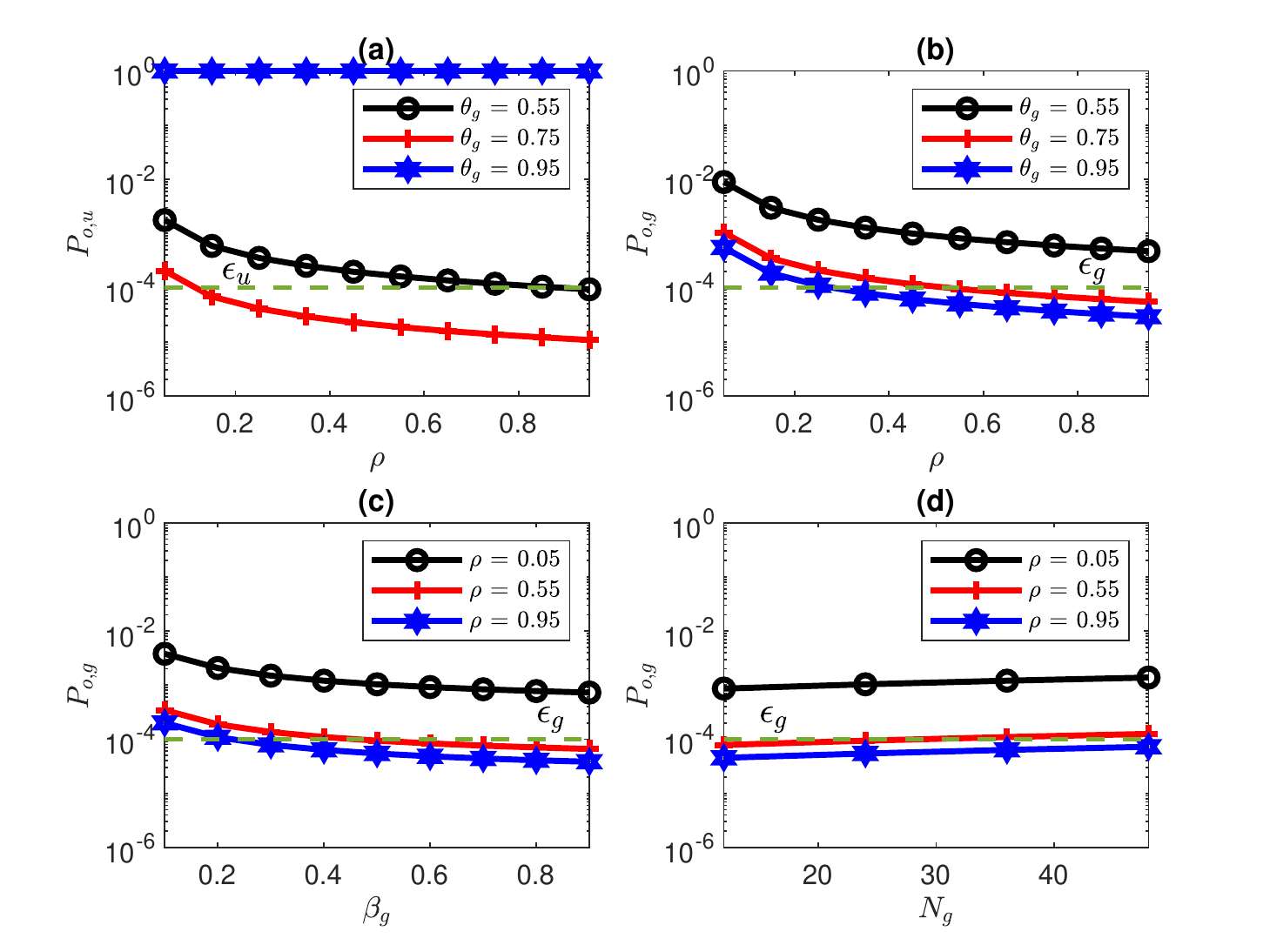}
		\caption{Outage probabilities: (a) $P_{o,u}$ vs. $\rho$, (b) $P_{o,g}$ vs. $\rho$, (c) $P_{o,g}$ vs. $\beta_g$ and (d) $P_{o,g}$ vs. $N_g$.}
		\label{outage_probs}
\end{figure}

\begin{figure}
		\centering
		\includegraphics[width=3.8in]{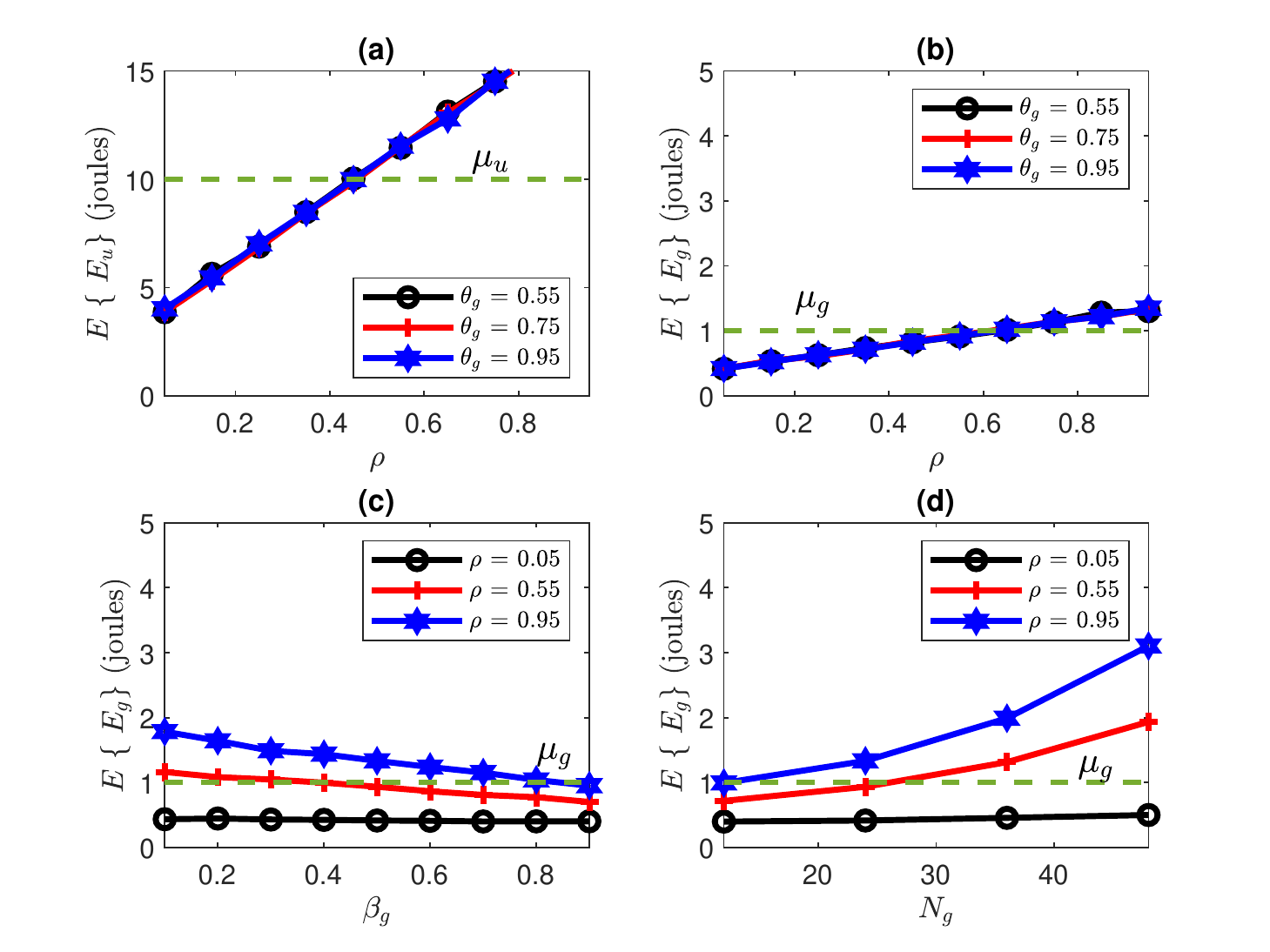}
		\caption{Average harvested energy: (a) $\mathbb{E}\{E_{u}\}$ vs. $\rho$, (b) $\mathbb{E}\{E_{g}\}$ vs. $\rho$, (c) $\mathbb{E}\{E_{g}\}$ vs. $\beta_g$ and (d) $\mathbb{E}\{E_{g}\}$ vs. $N_g$.}
		\label{harvested_energy}
\end{figure}

Fig. \ref{sec_rates} presents the simulated average secrecy rates. Fig. \ref{outage_probs} presents the simulated  outage probabilities. Fig. \ref{harvested_energy} presents the simulated harvested energy. In Figs. \ref{sec_rates} to \ref{harvested_energy}, we have (a) average parameter for $u$ versus $\rho$ for different $\theta_g$ values (b) average parameter for $g$ versus $\rho$ for different $\theta_g$ values, $N_g = 24$ and $\beta_g= 0.5$ (c) average parameter for $g$ versus $\beta_g$ for different $\rho$ values, $N_g = 24$ and $\theta_g= 0.75$ and (d) average parameter for $g$ versus $N_g$ for different $\rho$ values, $\theta_g = 0.75$ and $\beta_g= 0.5$. Note that changing $\theta_g$ changes $\theta_u$ since $\theta_u = 1-\theta_g$.

As shown in Fig. \ref{sec_rates} (a) as $\rho$ increases or $\theta_g$ decreases, the SNR at $u$ increases and hence $\mathbb{E}\{R_{s,u}\}$ increases. However since $g$ considers $u$'s signal as noise, as $\rho$ increases, the noise variance increases and hence $\mathbb{E}\{R_{s,g}\}$ decreases as shown in Fig. \ref{sec_rates} (b). Unlike $u$ as $\theta_g$ increases, the desired signal power at $g$ increases, which improves SNR and consequently $\mathbb{E}\{R_{s,g}\}$. As shown in Fig. \ref{sec_rates} (c) as $\beta_g$ increases, higher power factor is allocated for signal decoding rather than energy harvesting, which improves SNR as follows from the definition of $\Lambda_g$ and, hence, improves $\mathbb{E}\{R_{s,g}\}$. $N_g$ does not have much impact on $\mathbb{E}\{R_{s,g}\}$. Note that the upper bound (UB) in (\ref{R_sum}) is satisfied.
One key finding in Fig. \ref{outage_probs} (a) is that very low values of $\theta_u$, for example $0.05$, i.e., when $\theta_g= 0.95$, might lead to $P_{o,u} =1$. To achieve a desired $\epsilon_u \leq 10^{-4}$, $\rho$ can be $\in[0.05:0.95]$ and $\theta_g \in [0.55 : 0.9]$. As can be inferred from Fig. \ref{outage_probs} (b) to (d) to achieve a desired $\epsilon_g \leq 10^{-4}$, $\rho$ can be $\in[0.2:0.95]$,  $\theta_g \in [0.55 : 0.9]$, $\beta_g \in [0.05 : 0.95]$, and $N_g \in[12:N_s]$.

As expected, the harvested energy at $u$ is much higher than that at the far User $g$. To achieve a desired $\mu_u \geq 10$ joules, $\rho$ should be $\in[0.5:0.95]$ as shown in Fig. \ref{harvested_energy} (a). Note that since the energy is harvested at $u$ before cancelling User $g$'s signal, changing $\theta_g$ does not have an impact on $\mathbb{E}\{E_u\}$. As can be seen in Fig. \ref{harvested_energy} (b) to (d), to achieve $\mu_g \geq 1$ joule, $\rho$ should be $\in[0.65:0.95]$, $\beta_g$ should be $\in[0.55:0.95]$ and $N_g$ should be $\in [16:N_s]$. $N_g$ has the most impact on the harvested energy at $g$.
Following the exhaustive grid-based search operation, the optimal results under our simulation parameters are $\rho = 0.75$, $\theta_g = 0.75$, $\beta_g = 0.85$ and $N_g = 32$. The reader can infer these values from and the explanatory plots in Figs. \ref{sec_rates} to \ref{harvested_energy}. The achieved maximized $\mathbb{E}\{R_{s,u}\} = 4 $ bits/sec/Hz and $\mathbb{E}\{R_{s,g}\} = 0.7$ bits/sec/Hz while meeting the constraints on $P_{o,u} \leq 10^{-4}$, $P_{o,g} \leq 10^{-4}$,  $\mathbb{E}\{E_{u}\} \geq 10$ joules and $\mathbb{E}\{E_{g}\} \geq 1$ joule.

%%%====================================================================================
%%%====================================================================================
%%%====================================================================================
%%%====================================================================================
%%%====================================================================================
%%%====================================================================================
%%%====================================================================================
%%%====================================================================================
%%%====================================================================================
%%%====================================================================================
	\section{Conclusion}
	We presented an AN aided physical-layer security scheme for energy harvesting nodes operating under OFDM-based non-cooperative NOMA scheme. We showed that, with optimal choice of design parameters, it is possible to secure the considered system while transferring energy to multiple nodes under minimum outage probability constraints at legitimate users. %NOMA power split factor, AN to signal power split factor and number of samples used for PS after CP seem to have the most impact on the performance of the system. 
	\bibliographystyle{IEEEtran}
	\bibliography{references_NOMA_Journal}
\end{document}